\documentclass[pdflatex,referee,sn-nature]{sn-jnl}

\usepackage{graphicx}%
\usepackage{multirow}%
\usepackage{amsmath,amssymb,amsfonts}%
\usepackage{amsthm}%
\usepackage{mathrsfs}%
\usepackage[title]{appendix}%
\usepackage{xcolor}%
\usepackage{textcomp}%
\usepackage{manyfoot}%
\usepackage{booktabs}%
\usepackage{algorithm}%
\usepackage{algorithmicx}%
\usepackage{algpseudocode}%
\usepackage{listings}%
\usepackage{fullpage}
\usepackage{lineno}

\newcommand{\zeroth}{0$^{\mathrm{th}}${}}
\newcommand{\first}{1$^{\mathrm{st}}${}}
\newcommand{\second}{2$^{\mathrm{nd}}${}}

\theoremstyle{thmstyleone}%
\newtheorem{theorem}{Theorem}%  meant for continuous numbers
\newtheorem{proposition}[theorem]{Proposition}% 

\theoremstyle{thmstyletwo}%
\newtheorem{example}{Example}%
\newtheorem{remark}{Remark}%

\theoremstyle{thmstylethree}%
\newtheorem{definition}{Definition}%

\begin{document}
% \linenumbers
\title[Article Title]{X-ray Strain and Stress Tensor Tomography}


\author*[1,2]{\fnm{Peter} \sur{Modregger}}\email{modregger@uni-siegen.de}

\author[3]{\fnm{James A. D.} \sur{Ball}}

\author[1,2]{\fnm{Felix} \sur{Wittwer}}

\author[1,2]{\fnm{Ahmar} \sur{Khaliq}}

\author[3]{\fnm{Jonathan} \sur{Wright}}

\affil[1]{\orgdiv{Physics Department}, \orgname{University of Siegen}, \orgaddress{\street{Walter-Flex-Str. 3}, \city{Siegen}, \postcode{57072}, \country{Germany}}}

\affil[2]{\orgdiv{Centre for X-ray and Nano Science CXNS}, \orgname{Deutsches
Elektronen-Synchrotron DESY}, \orgaddress{\street{Notkestr. 86}, \city{Hamburg}, \postcode{22607},  \country{Germany}}}

\affil[3]{\orgdiv{ID11}, \orgname{European Synchrotron Radiation Facility ESRF}, \orgaddress{\street{71 Av. des Martyrs}, \city{Grenoble}, \postcode{38000},  \country{France}}}

%\affil[3]{\orgdiv{Department}, \orgname{Organization}, \orgaddress{\street{Street}, \city{City}, %\postcode{610101}, \state{State}, \country{Country}}}


\abstract{The microscopic distribution of strain and stress plays a crucial role for the performance, safety, and lifetime of components in aeronautics, automotive and critical infrastructure~\cite{Schijve2009}. While non-destructive methods for measuring the stress close to the surface have long been long established, only a limited number of approaches for depth-resolved measurements based on x-rays or neutrons are  available~\cite{Fitzpatrick2003}. These feature significant limitations, including long scan times, intricate experimental set-ups, limited spatial resolution or anisotropic gauge volumes with aspect ratios of 1:10 or worse. Here, we present a method that overcomes these limitations and obtains tomographic reconstructions of the full six-dimensional strain and stress tensor components. Using a simple and wide spread experimental set-up that combines x-ray powder diffraction with single axis tomography, we achieve non-destructive determination of depth-resolved strain and stress distributions with isotropic resolution. The presented method could be of interest for additive manufacturing of metals~\cite{Herzog2016,Uzun2024}, battery research~\cite{Li2022}, in-situ metallurgy~\cite{Isavand2021} and the experimental validation of finite element simulations~\cite{Schoinochoritis2017}.}

\keywords{x-ray diffraction, tomography, stress tensor, non-destructive measurement}



\maketitle

% \section{Introduction}\label{sec1}

\begin{figure}[htbp]
    \centering
    \includegraphics[width=\linewidth]{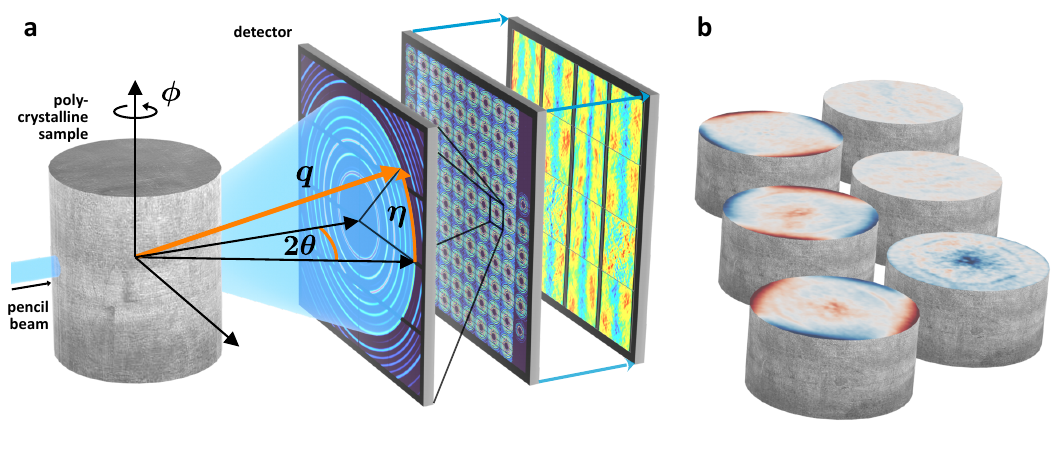}
    \caption{Experimental set-up, data analysis and stress tensor maps. (a) Schematic representation of the experimental set-up. The sample is rotated and  scanned horizontally through an hard x-ray pencil beam and powder diffraction patterns are acquired at each scan position. These are then analysed with respect to the intensity and peak position in 8 angular segments and over 2 diffraction peaks resulting into 16 sinograms as shown. (c) Maps of the six reconstructed stress tensor components visualized as cuts through the sample.}
    \label{fig:STT_setup}
\end{figure}

The possibility of combining x-ray powder diffraction with tomography for the depth-resolved measurement of the strain tensor field $\epsilon_{ij}(x,y)$, given by 
\begin{equation}
    \epsilon_{ij}(x,y) = 
    \left( 
    \begin{array}{ccc}
    \epsilon_{11}(x,y) & \epsilon_{12}(x,y) & \epsilon_{13}(x,y)\\
    \epsilon_{12}(x,y) & \epsilon_{22}(x,y) & \epsilon_{23}(x,y)\\
    \epsilon_{13}(x,y) & \epsilon_{23}(x,y) & \epsilon_{33}(x,y)\\
     \end{array}
    \right),    
\end{equation}
has been discussed for almost two decades~\cite{Korsunsky2006, Jun2010, Lionheart2015}. The corresponding experimental set-up of this approach, which we would like to call x-ray strain/stress tensor tomography (STT), features a monochromatic pencil beam provided by a hard x-ray beamline at a synchrotron radiation facility, a poly-crystalline sample positioned on several translation axes and one rotation axis as well as an area detector for collecting the diffraction signal (see Fig.~\ref{fig:STT_setup}a).

The key challenge resides in the anisotropic contribution of the projected strain field during tomographic rotation around the angle $\phi$. The observable angular shift due to strain of the diffraction curves in terms of the \first{} angular moment ($M_1$) is given by the differential Bragg equation~\cite{Hauk1997} and the projection of the strain tensor onto the scattering vector $\bf q$
\begin{equation}\label{eq:strain_projection}
    M_1(x,y) = - \tan\theta \sum_{ij} q_i q_j \epsilon_{ij}(x,y)
\end{equation}
with ${\bf q} = (q_1, q_2, q_3)$, the components of the scattering vector. Since the scattering vector changes orientation during tomographic rotation (i.e., $\bf q = \bf q(\phi)$), the observable strain varies accordingly. In fact, STT shares this challenge with related prominent techniques such as X-ray tensor tomography based on dark-field imaging~\cite{Malecki2014,Kim2020}, x-ray scattering tensor tomography~\cite{Stock2008,Liebi2015,Schaff2015} and x-ray texture tomography~\cite{Frewein2024,Carlsen2025}. These techniques usually use two rotation axes in order to retrieve the local orientation of intensity signals. In 2015, Lionheart and Withers have mathematically proven that line scans around six carefully chosen rotation axes and data from one diffraction ring provides sufficient information for the tomographic reconstruction of the six strain tensor components~\cite{Lionheart2015}. However, they explicitly leave open the possibility for using fewer rotation axes. Here, we will use the powder diffraction information from eight angular segments of two diffraction rings. Effectively, the \second diffraction ring acts as a \second rotation axis (i.e., also $\bf q = \bf q(\theta)$), which avoids an under-determination of the inversion problem. 

The challenge of anisotropic strain contributions during rotation is significantly alleviated when the diffraction signal can be traced back to specific grain positions in the sample. Methods taking advantage of this, such as diffraction contrast tomography~\cite{King2008,Ludwig2009} or (scanning) 3D x-ray diffraction~\cite{Poulsen2001,Henningsson2020}, allow for the determination of strains within single grains. Obviously, these techniques rely on the separability of diffraction peaks originating from different grains. Peak overlap constitutes a substantial complication~\cite{Henningsson2024}, which limits the applicability to comparatively large grains. For STT, on the other hand, we assume the opposite. Within each voxel the integrated diffraction intensity should ideally be isotropic with respect to tomographic rotation. In terms of the \zeroth angular moment this implies $M_0(x,y,\phi) \approx M_0(x,y)$. Thus, STT works best for numerous grains in the order of hundreds in each voxel corresponding to an absence of crystallographic texture.

% Ludwig W, King A, Reischig P, Herbig M, Lauridsen EM, Schmidt S, et al. New opportunities for 3D materials science of polycrystalline materials at the micrometre length scale by combined use of X-ray diffraction and X-ray imaging. Mater Sci Eng A 2009;524:69–76.
% THIS IS A RECAP OF DCT

% Poulsen 2005 - Amorphous materials; needs 0 external force for reference and external force to have measurements. Fit one ring to obtain three planar epsilons. Map was in 2d projection.
% \red{Measuring strain distributions in amorphous materials HENNING F. POULSEN1\cite{Poulsen2005}} THIS IS JUST 2D

\begin{figure}[htbp]
    \centering
    \includegraphics[width=0.5\linewidth]{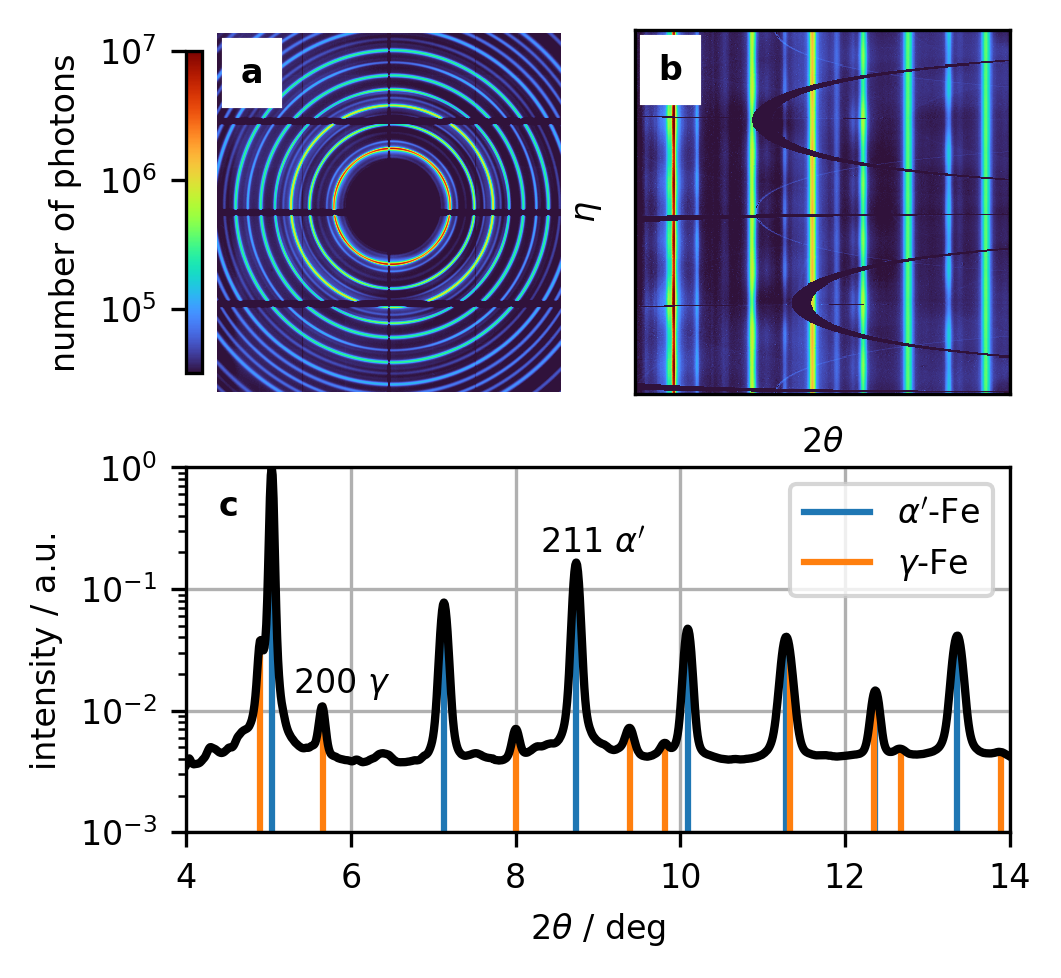}
    \caption{Average diffraction from the martensite sample. (a) Summed diffraction pattern, which is also shown in Fig.~\ref{fig:STT_setup}a. (b) Caked diffraction pattern. (c) Azimuthal diffraction pattern revealing that the sample consists of mostly martensite ($\alpha'$-Fe) with some retained austenite ($\gamma$-Fe).
    }
    % This is 06_rod strain ringinds[2,6] lambda35 Niter 222 13306sec savinteger82617 
    \label{fig:avg_diffpatt}
\end{figure}

\begin{figure}
    \centering
    \includegraphics[width=0.5\linewidth]{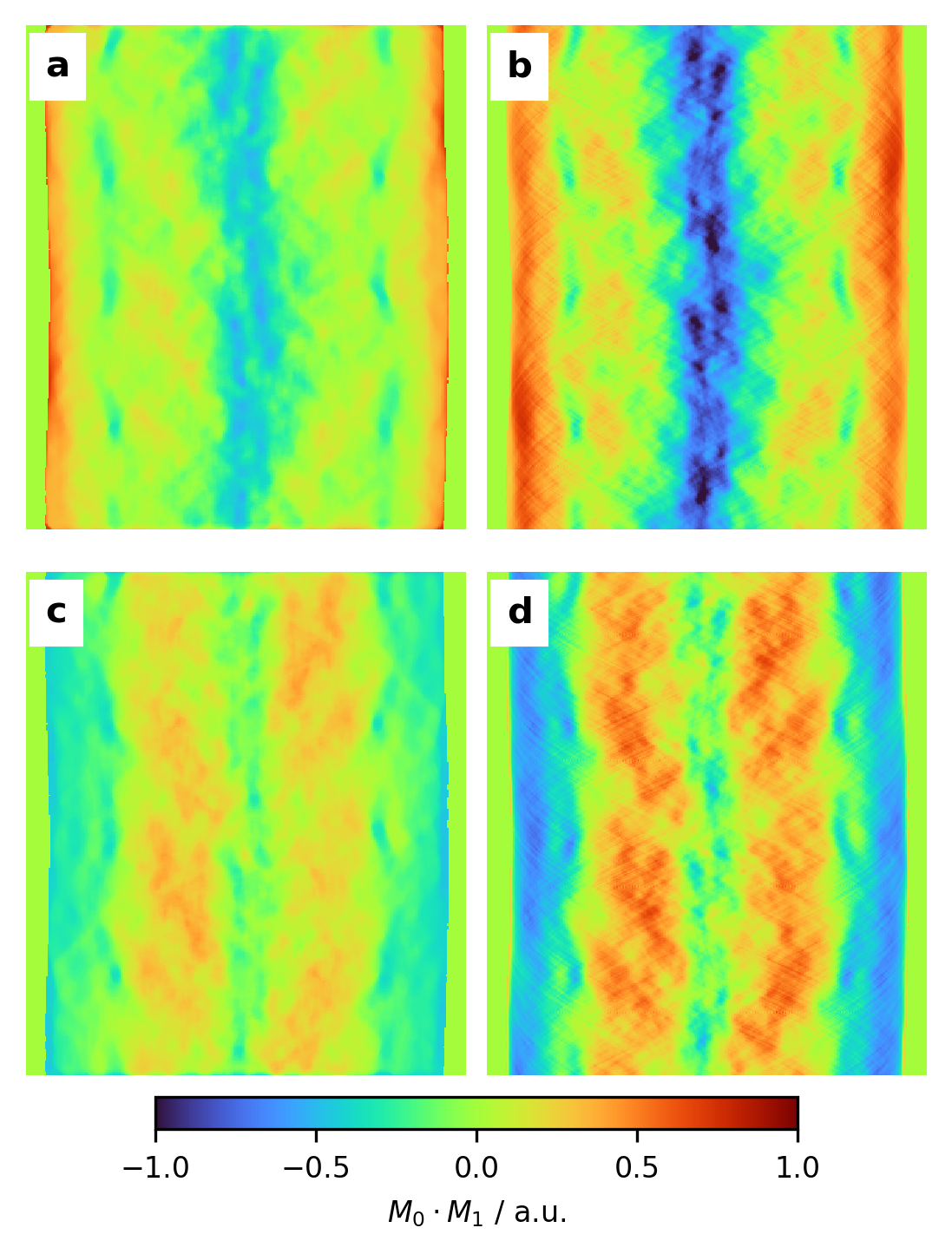}
    \caption{Comparison between experimental and retrieved un-normalized $M_0\cdot M_1$ sinograms. (a) and (c)  are the experimental sinograms for $\eta =-180$\textdegree{} and $\eta=0$\textdegree{} of the 211 peak, which have been corrected for parallax effects~\cite{Modregger2024a}. (b) and (d) are the corresponding retrieved sinograms. Correlation coefficients between experimental and retrieved sinograms are $r=0.77$ and $r=0.66$, respectively.}
    \label{fig:exp_vs_model}
\end{figure}

The experimental scan consists of tomographic rotation around $\phi$ as the outer and transversal translation $t$ as the inner loop. Scans take a few hours, where in the order of 100,000 diffraction patterns are collected (see Extended Data Figure~\ref{fig:avg_diffpatt}). Noticeably, parallax at the detector, i.e. the apparent lateral offset of diffraction occurring at different sample depths, cancels out exactly if the rotation is performed over 360\textdegree{}~\cite{Modregger2024a}. Collected diffraction patterns are integrated in the azimuthal direction with pyFAI~\cite{Kieffer2020} over eight even spaced angular segment and with two diffraction rings. The \zeroth moment, $M_0$, and the \first moment, $M_1$ of the resulting diffraction curves are calculated~\cite{Frederik} and used as input for iterative minimization of the cost function
\begin{equation}\label{eq:cost_function1}
    L = \sum_{t,\phi,\theta,\eta} \left({M}_0  M_1 - \sum_{ij}\mathcal{P}_{\phi\theta\eta}\left[\mathcal{R}^{-1}[M_0] \cdot \epsilon_{ij} \right]   \right)^2 + \lambda \sum_{ij} \mathrm{TV}[\epsilon_{ij}].
\end{equation}
Here, $\mathcal{P}_{\phi\theta\eta}$ is the projection operator and $\mathcal{R}^{-1}$ the inverse Radon transform~\cite{Kak1988}. Compared to~\cite{Korsunsky2011,Uzun2024} one of the key insight here is that the un-normalized \first moments (i.e., $M_0\cdot M_1$ with the latter from eq.~\ref{eq:strain_projection}) have a well-defined Radon transform, while the normalized \first moments do not. For noise suppression total energy variation $\mathrm{TV}$ was included with $\lambda$ as user-chosen strength factor. Retrieved strain tensor components constitute the average over utilized reflections and over crystallites within one voxel. The experimental sinograms were in good agreement with their model prediction (see Extended data Figure~\ref{fig:exp_vs_model}). Further, strain reconstruction has been repeated for three pairs of diffraction rings (i.e., two out of 211, 310 and 321; see Extended Data Fig.~\ref{}) and high correlation coefficients ranging from $r=0.832$ to $r=0.985$ have been observed. Courteously, sensitivities of the \first angular moment are in the range of $0.1\,\mu$rad~\cite{Modregger2025a}, which results in expected strain sensitivities below $10^{-4}$.

\begin{figure}
    \centering
    \includegraphics[width=0.5\textwidth]{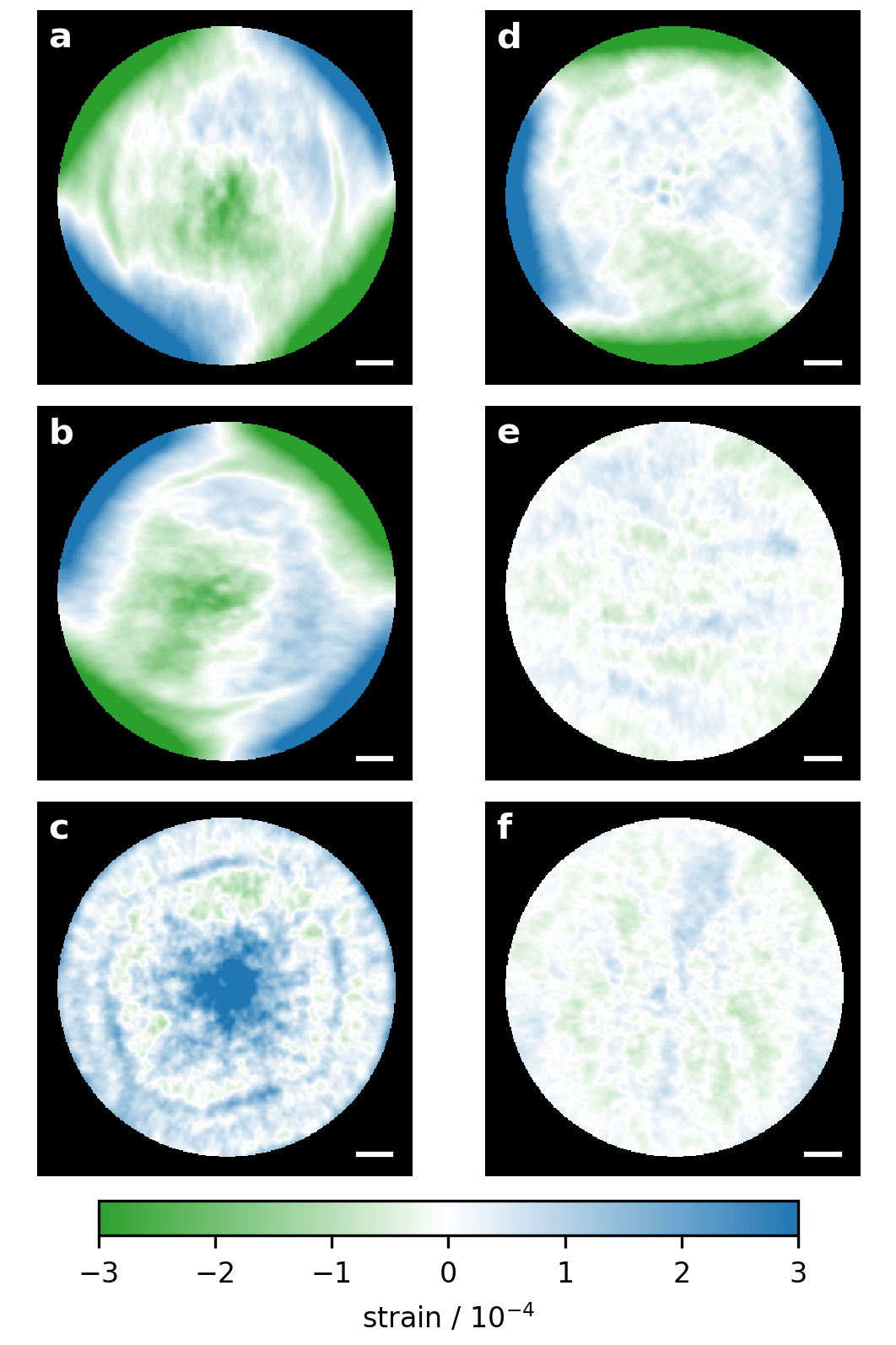}
    \caption{Maps of reconstructed strain tensor components of a spring steel rod with residual strains. (a) $\epsilon_{11}$, (b) $\epsilon_{22}$, (c) $\epsilon_{33}$, (d) $\epsilon_{12}$,
    (e) $\epsilon_{13}$, (f) $\epsilon_{23}$. Scale bars are 300~$\mu$m.}
    % This is 06_rod strain ringinds[2,6] lambda35 Niter 222 13306sec savinteger82617 
    \label{fig:control_strains}
\end{figure}

\begin{figure}[htbp]
    \centering
    \includegraphics[width=0.5\linewidth]{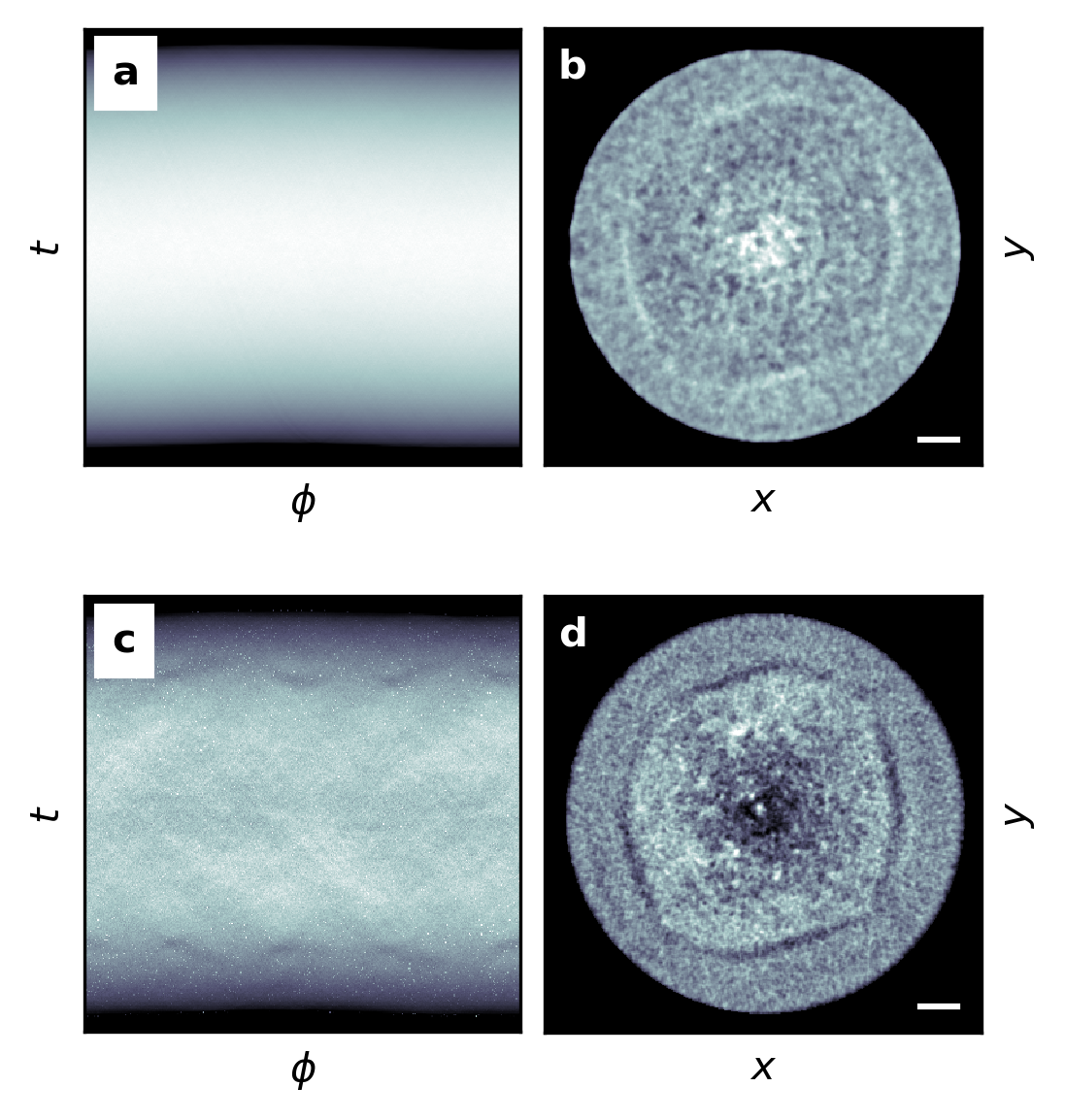}
    \caption{Integrated intensity sinograms and tomographic reconstruction of the two peaks indicated in~Fig.~\ref{fig:avg_diffpatt}c. (a) $M_0$ sinogram of the martensite 211 peak and (b) its reconstruction. (c) $M_0$ sinogram of the austenite 200 peak and (d) its reconstruction. Both sinograms show a clear absence of crystallographic texture on the scale of the utilized beamsize.}
    \label{fig:M0sinoM0rec}
\end{figure}

\begin{figure}[htbp]
    \centering
    \includegraphics[width=0.5\textwidth]{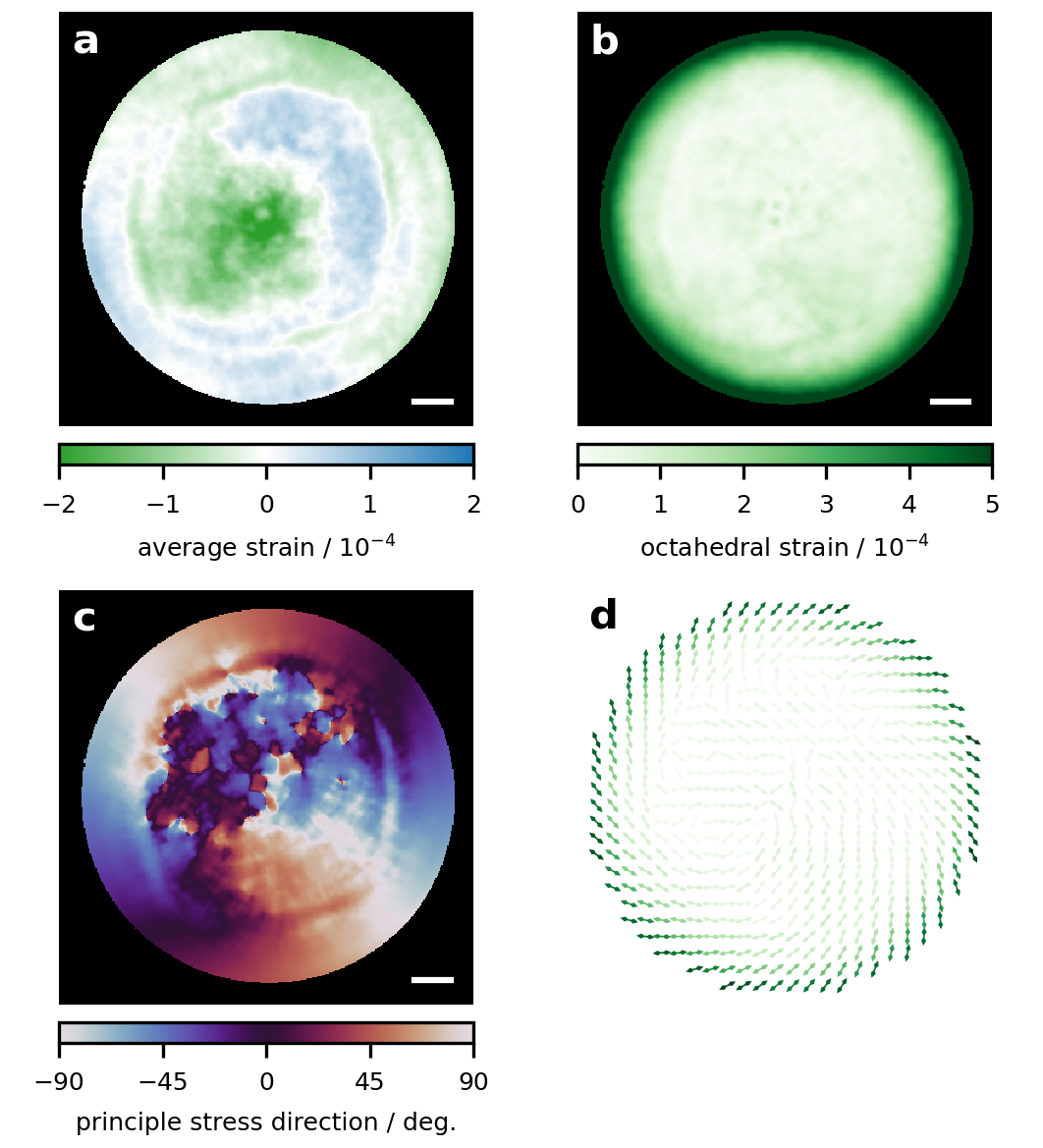}
    \caption{Principle strain tensor components of the spring steel rod. (a) Average strain, (b) octahedral strain, (c) principle strain direction, (d) Visualization of the principle strain direction colored by the octahedral strain. Scale bars are 300~$\mu$m.}
    % This is 06_rod strain ringinds[2,6] lambda35 Niter 222 13306sec savinteger82617 
    \label{fig:control_principle_strains}
\end{figure}

Using the STT technique described above, we reconstructed the full 6D strain tensor within each voxel in a tomographic slice of a martensitic spring steel rod with a diameter of 3.2~mm (see Fig.~\ref{fig:control_strains}). The observation of small residual strains was expected since these are thermally induced during uneven cooling of the component (outer parts cools down faster than inner parts). However, we have observed unexpected large in-plane strains (i.e. $\epsilon_{11}, \epsilon_{22},\epsilon_{12}$) with corresponding hydrostatic strain values close to zero (see Extended Data Fig.~\ref{fig:control_principle_strains}). The out-of-plane components, on the other hand, follow expectations. Shear strains $\epsilon_{13}$ (Fig.~\ref{fig:control_strains}e) and $\epsilon_{23}$ (Fig.~\ref{fig:control_strains}f) are vanishingly small, but can be used to determine the achieved strain sensitivity by their standard deviation resulting in $u(\epsilon) \approx 3\cdot10^{-5}$. The normal out-of-plane component $\epsilon_{33}$, shown in Fig.~\ref{fig:control_strains}c, exhibits the expected thermally induced behaviour: A strong tensile strain in the centre that tapers of towards the edge and locally changes to compressive strain. The radial symmetry is broken by a deformed ring at about 400~$\mu$m from the edge that reveals the orientation of the original square shape of the component prior to wire cold-working. This deformed ring is due to macro-segregation of impurities during cooling~\cite{Hunkel2021}, contains comparatively little retained austenite (see Extended Data Fig.~\ref{fig:M0sinoM0rec}) and exhibits compressive in-plane but tensile out-of-plane strain. 

\begin{figure}
    \centering
    \includegraphics[width=0.5\textwidth]{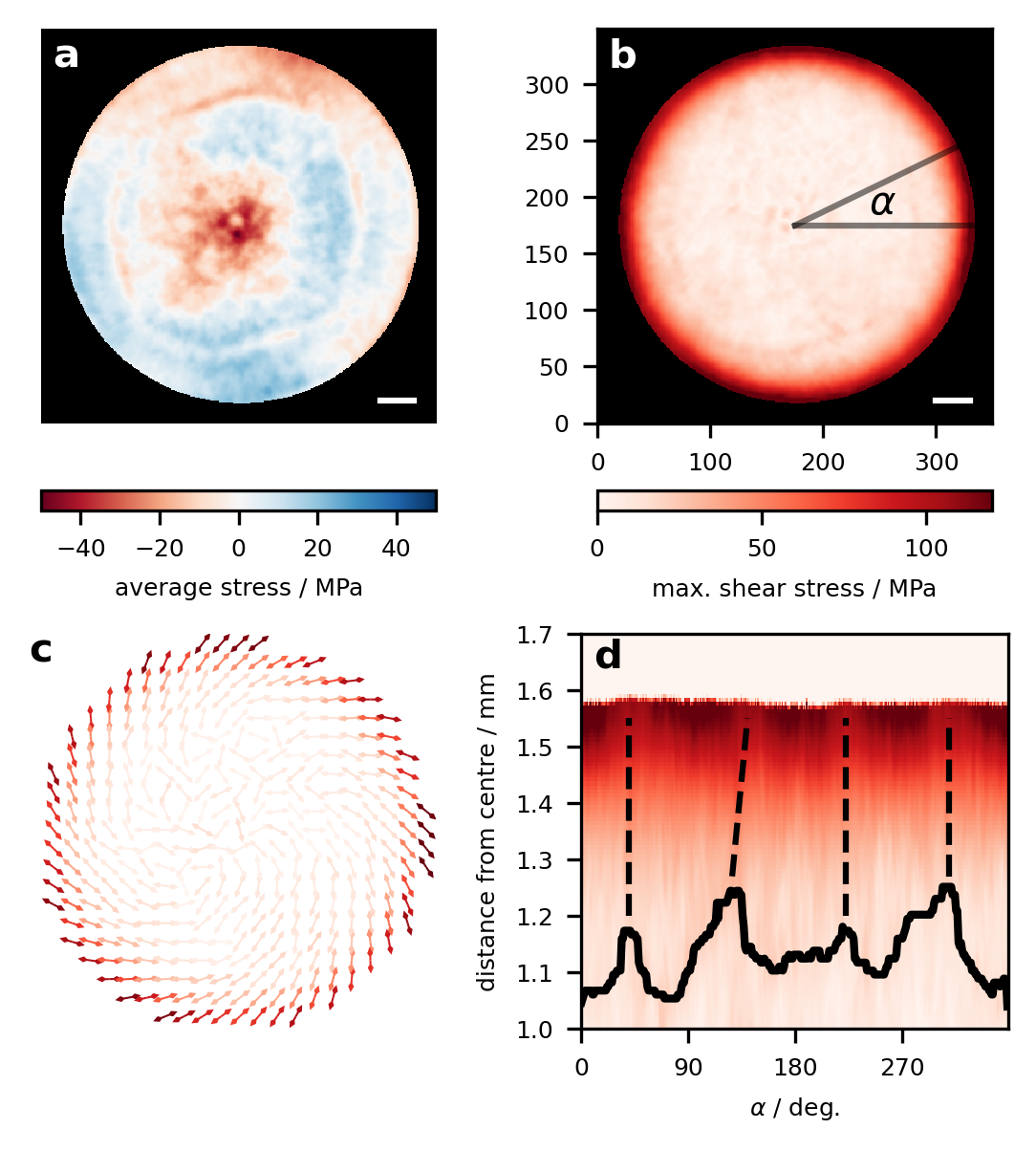}
    \caption{In-plane principle stresses of a spring steel rod. (a) Average stress and (b) maximum shear stress. (c) The visualization of the principle stress direction reveals a torsional stress state with clock-wise orientation. (d) Polar representation of maximum shear stresses (b) with the position of the segregation ring indicated by the black line. Dashed lines indicate an anti-correlation between the ring distance from surface and stress values at the surface. The colour bar in (b) also applies to (c) and (d). Scale bars are 300~$\mu$m.}
    % This is 06_rod stress ringinds[2,6] lambda35 Niter 219 13095sec savinteger26380
    \label{fig:principle_stresses}
\end{figure}

Usually, strain is introduced as the reaction of a body to an applied stress $\sigma$ and the well known Hooke's law $\sigma = E\epsilon$~\cite{hooke2016lectures} connects those two via Young's modulus $E$. Naturally, it is the stress that determines possible material failure \cite{Schijve2009} and the knowledge of its internal distribution is essential for safety evaluations and lifetime predictions. Using the generalized Hooke's law, which connects strain to stress tensors~\cite{Noyan1987}, and taking into account the elastic anisotropy of the material via so-called x-ray elastic constants~\cite{Manns2013}, the cost function eq.~(\ref{eq:cost_function1}) can simply be adapted for the reconstruction of stress tensor components $\sigma_{ij}$. Retrieved stress values constitute the average over grains in a voxel, thus macro residual stresses are reconstructed.

Force balance requires the in-plane stresses (i.e., $\sigma_{11}$ and $\sigma_{22}$) to sum to zero, since the sample would move or deform on its own otherwise. Here, we have used this physical constraint in order to determine the strain-free lattice parameter required for the determination of strain. For this reason, stress tensor reconstruction was performed prior to strain tensor reconstruction. Retrieved stress tensor components are shown in the Extended Data Figure~\ref{fig:principle_stresses}, which suggest an achieved stress sensitivity of about 20~MPa. Principle stresses, i.e., the average (hydrostatic) stress, the maximum shear stress and the principle stress direction are shown in Fig.~\ref{fig:principle_stresses}. Principle strains are provided in Extended Data Figure~\ref{fig:control_principle_strains}. Restricting the discussion to the dominant in-plane direction, we have observed a close to vanishing hydrostatic stress (Fig.~\ref{fig:principle_stresses}a) and a maximum shear stress that increases from the segregation ring to the edge of the sample (Fig.~\ref{fig:principle_stresses}b). The principle stress direction, shown in Fig.~\ref{fig:principle_stresses}c reveals an unexpected and -- to the supplier unknown -- torsional stress state with clock-wise orientation. This already constitutes a first application example of STT. Originally, the steel rod was supposed to be manufactured into a spring and the winding direction would have been selected assuming no residual shear stresses. However, in the presence of the torsional stress the choice of the winding direction will either significantly increase or decrease the lifetime of the spring. Furthermore, Fig.~\ref{fig:principle_stresses}d shows presentation of the maximum shear stresses in polar coordinates. Here, we found a clear anti-correlation between the shear stress values at the sample surface and the distance to the deformed segregation ring.

% \red{LAB results for stress are good with respect to stress values at surface}

%\begin{figure}
%    \centering
%    \includegraphics[width=0.45\linewidth]%{stresses_shotpeened.png}
%    \includegraphics[width=0.45\linewidth]{uneven_shotpeening.png}
%    \caption{\red{LEFT: Stresses from the shot peened samples. Not sure if I should keep them in (if then of course as better versions). However, the reconstruction only almost works. RIGHT: Uneven coverage of shot-peening revealed by sigma33. This was already interesting for the shot peening guys. However, technically speaking one doesnt need STT for this, as the old Korsunsky method also delivers at least the corresponding strain. This would be similar to Fig.~\ref{fig:korsunsky}d. Maybe the difference between strain and stress is enough here?  }}
%    \label{fig:stresses_shotpeened}
%\end{figure}

\begin{figure}[htbp]
    \centering
    \includegraphics[width=0.5\linewidth]{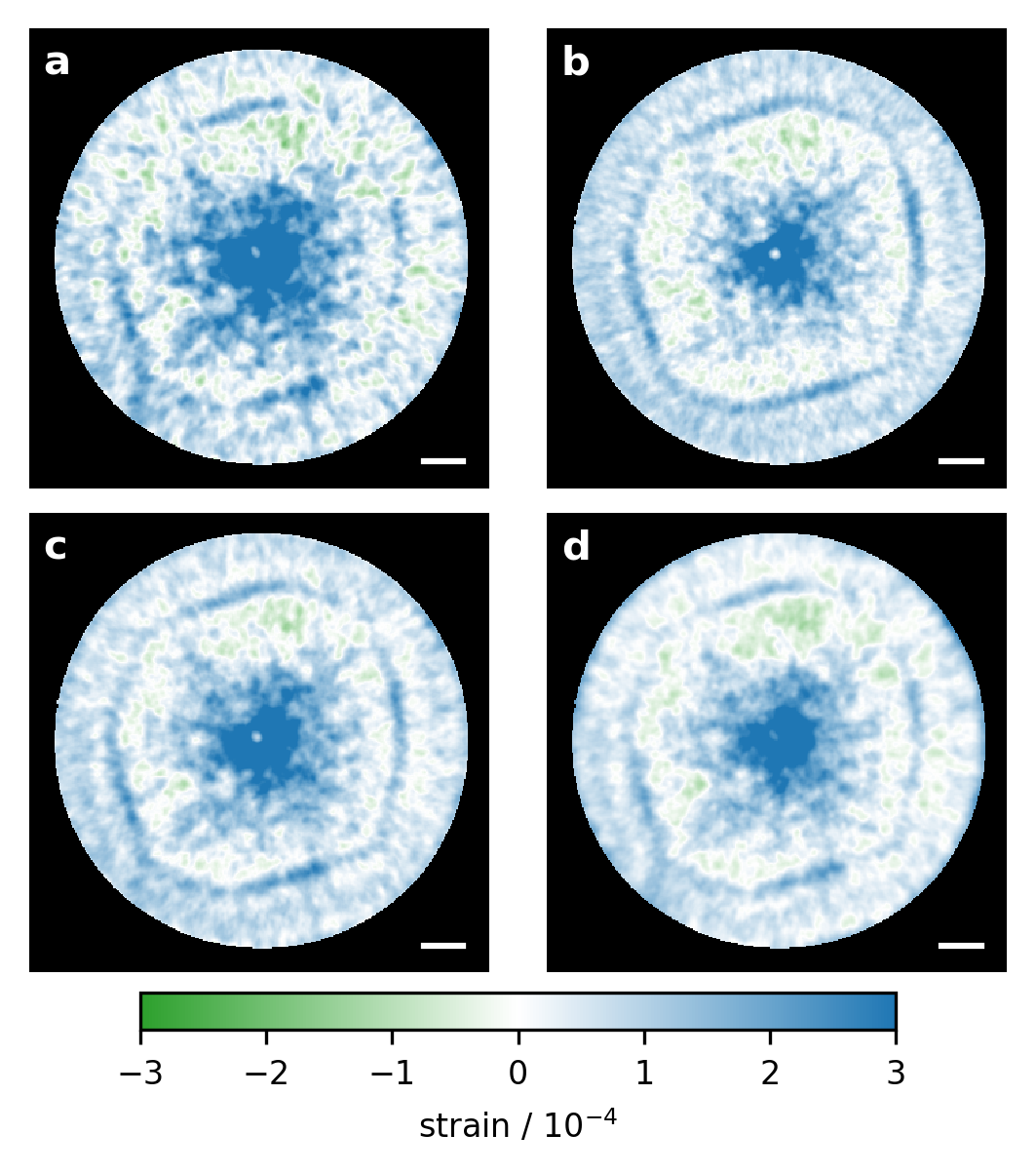}
    \caption{Comparison of $\epsilon_{33}$ reconstruction between the approach by Korsunsky et al.~\cite{Korsunsky2011} and the proposed approach. (a) Korsunsky reconstruction from the 211 peak. (b) Korsunsky reconstruction from the 321 peak. (c) The average of (a) and (b). (d) $\epsilon_{33}$ retrieved by the proposed approach (same as Fig.~\ref{fig:control_strains}c repeated here for convenience). Correlation coefficients between the proposed method and the other reconstructions are $r=0.94$, $r=0.82$ and $r=0.96$, respectively.}
    % This is 06_rod strain ringinds[2,6] lambda35 Niter 222 13306sec savinteger82617 
    \label{fig:korsunsky}
\end{figure}

In conclusion, we have demonstrated the simultaneous tomographic reconstruction of all six strain and all six stress tensor components in a martensitic steel rod cross-section (i.e., 8D data). Compared to established x-ray based methods for depth-resolved stress mapping such as conical slit cells ~\cite{Staron2014} or energy dispersive x-ray diffraction~\cite{Apel2018} the experimental set-up is simple featuring only a pencil beam, a single rotation axis and an x-ray area detector providing an isotropic gauge volume. These minimal requirements render the presented approach ready to use for in-situ measurements as well as compatible with a large number of existing beamlines at synchrotron radiation facilities indicating a wide spread use. Unlike diffraction strain tomography, where the out-of-plane tensor components are experimentally determined~\cite{Korsunsky2011} and the other components may be retrieved by eigenstrain reconstruction if several experimental slices are available~\cite{Uzun2024}, here, all tensor components are directly measured. Intriguingly, eigenstrain reconstruction should greatly benefit from the availability of the six tensor components provided by STT. In fact, we found a very good agreement between the out-of-plane stress tensor component reconstructed by diffraction strain tomography and STT (see Extended Data Fig.~\ref{fig:korsunsky}).

Due to the absence of noticeable crystallographic texture, the sample used here was especially suitable for STT. The challenge of imaging materials systems with larger grain sizes may be met in two ways. First, the beamsize can be increased in order to collect the diffraction signal of more grains. We estimate that grain sizes of 10~$\mu$m produce a texture free diffraction signal with a beamsize of 1~mm. Second and more exciting, is the possible combination of STT with texture tomography, where the latter was demonstrated to be compatible with a single rotation axis~\cite{Carlsen2025}.

Nevertheless, we believe that the non-destructive and depth-resolved measurement of 8D elastic tensor properties in powder-like samples is of great interest to materials science research including the impact of stress on the life time of batteries and additive manufactured metals. Grain mapping techniques such as scanning 3D x-ray diffraction cover intra-granular stresses (type III), while STT covers macro stresses (type I). It now stands to reason that inter-granular stresses (type II) become fully accessible by improving the tolerance of grain mapping for peak overlapping and/or the tolerance of STT for crystallographic texture.

\section*{Methods}

\subsubsection*{Sample}

The sample was an oil-tempered FDSiCr spring steel rod (EN 10270-2 FD SiCr) with a diameter of 3.2~mm supplied by Federnwerke J.P.~Grueber (Hagen, Germany). The manufacturing process started with a cold-worked wire rod with a diameter of 6~mm, which was reduced to 3~mm by a fine wire drawing machine. Subsequent recrystallization annealing, tempering and quenching produced the final martensite composition. Prior-austenite grains were micrometer in size, which lead to a nearly absence of crystallographic texture in the x-ray diffraction experiment. Retained austenite content was about 5 wt\% as determined by standard Rietveld refinement.

\subsubsection*{Experiment}

Synchrotron radiation-based experiments were carried out at the nanofocus station of the ID11 beamline at the ESRF in Grenoble, France~\cite{Wright2020}. A monochromatic beam with a photon energy of 69~keV was provided by a horizontally oriented Si (111) double Laue monochromator. The x-ray beam was collimated by 32 Al lenses shaped by perpendicular slits to a size 10~$\mu$m by 10~$\mu$m. A photon counting Eiger2 X CdTe 4M detector (Dectris, Switzerland) approximately 0.3~m downstream of the sample was used to collect the diffraction signals. Calibration of the set-up geometry was performed with a CeO$_2$ calibrant powder and pyFAI~\cite{Kieffer2020}. Tomographic scans involved continuous rotation of the sample over 360\textdegree{} in 400 steps as the inner loop and lateral translation by $10$~$\mu$m with 350 steps as the outer loop. Diffraction patterns were acquired with an exposure time of 50~ms resulting in a total scan time of 2.5~h for 140,000 diffraction patterns including overhead. 

\subsubsection*{Reconstruction}

Each diffraction pattern was integrated in the azimuthal direction over 8 angular segments with centres at $\eta$ = -180\textdegree{}, -135\textdegree{}, -90\textdegree{}, -45\textdegree{}, 0\textdegree{}, 45\textdegree{}, 90\textdegree{} and 135\textdegree{}. The 211 and 321 peaks were selected for further processing due to their high multiplicity. The resulting diffraction peaks $I(\Delta\theta)$ were background corrected and their \zeroth moment $M_0=\sum I(\Delta\theta)$ and \first moment $M_1=\sum \Delta\theta\, I(\Delta\theta) / M_0$ calculated. Similar to x-ray diffraction tomography~\cite{Bleuet2008}, the sinogram of the \zeroth moment was tomographically reconstructed yielding $\mathcal{R}^{-1}[M_0]$, which was corrected for absorption with the transmission sinogram.

Strain and stress tensor components were retrieved by iterative minimization of the cost function~eq.(\ref{eq:cost_function1}) using the limited memory implementation of the Broyden--Fletcher--Goldfarb--Shannon~\cite{Byrd1995a} algorithm in the SciPy library for Python~\cite{Jones2001}. The scattering vector $\bf q$ in the utilized coordinate system was
\begin{equation}
    {\bf q} = 
    \begin{pmatrix} 
    \cos\phi \cos\theta \cos\eta + \sin\theta \sin\phi \\ 
    \sin\phi \cos\theta \cos\eta - \sin\theta \cos\phi \\
    - \cos\theta \sin\eta
    \end{pmatrix}.
\end{equation}
Typical reconstruction took around 200 iteration steps taking approximately 4~h on a standard PC, where available analytical gradients sped up the iteration by orders of magnitude. The noise suppression factor was $\lambda = 35$ for all reconstructions. For stress tensor reconstructions X-ray elastic constants (i.e., $s_1(hkl)$ and $s_2(hkl)/2$) of steel have been determined by the DECcalc software~\cite{Manns2013} and taken into account by multiplying the projector $P_{\phi\theta\eta}$ with
\begin{equation}
    m_{hkl} = -\frac{s_1(hkl)}{\nu / E} \approx  \frac{\frac{s_2}{2}(hkl)}{ \frac{1+\nu}{E}},
\end{equation}
$\nu$ is Poisson's ration and the multipliers $m_{hkl}$ represent an $hkl$-dependent deviation from bulk properties between measured strain and actual stress. 

\section*{Data availability}
The raw data used to produce the figures of this paper are available at
https://doi.org/10.15151/ESRF-ES-1729370923. 

\section*{Code availability}
The code for stress tensor tomography is available via PM upon reasonable request.

\section*{Acknowledgments}
We acknowledge the European Synchrotron Radiation Facility (ESRF) for provision of synchrotron radiation facilities under proposal number MI-1498 and we would like to thank the beamline staff for assistance and support in using beamline ID-11. Parts of these investigations were funded by the ErUM-Pro programme (grant number 05K22PS2) of the German Federal Ministry of Education and Research (BMBF). This research was supported in part through the Maxwell computational resources operated at DESY.

\section*{Contributions}
P.M. conceived and designed the research, implemented the numerical code, analysed the data and wrote the paper. J.B. conducted the experiment, analysed the data and edited the paper, F.W., A.K. and J.W. conducted the experiment and edited the paper.

\section*{Competing interests}
The authors declare no competing interests.

\bibliography{stt_nature}

\newpage
\section*{Extended Data Figures}

%%%% Average diffration pattern %%%%%

%%%% M0sinoM0rec %%%%%

% L vs Niter

%%%% EXP vs MODEL %%%%%

%%%% principle strain %%%%%

%%%% comparison between peak pairs %%%%%
\begin{figure}[htbp]
    \centering
    \includegraphics[width=0.5\linewidth]{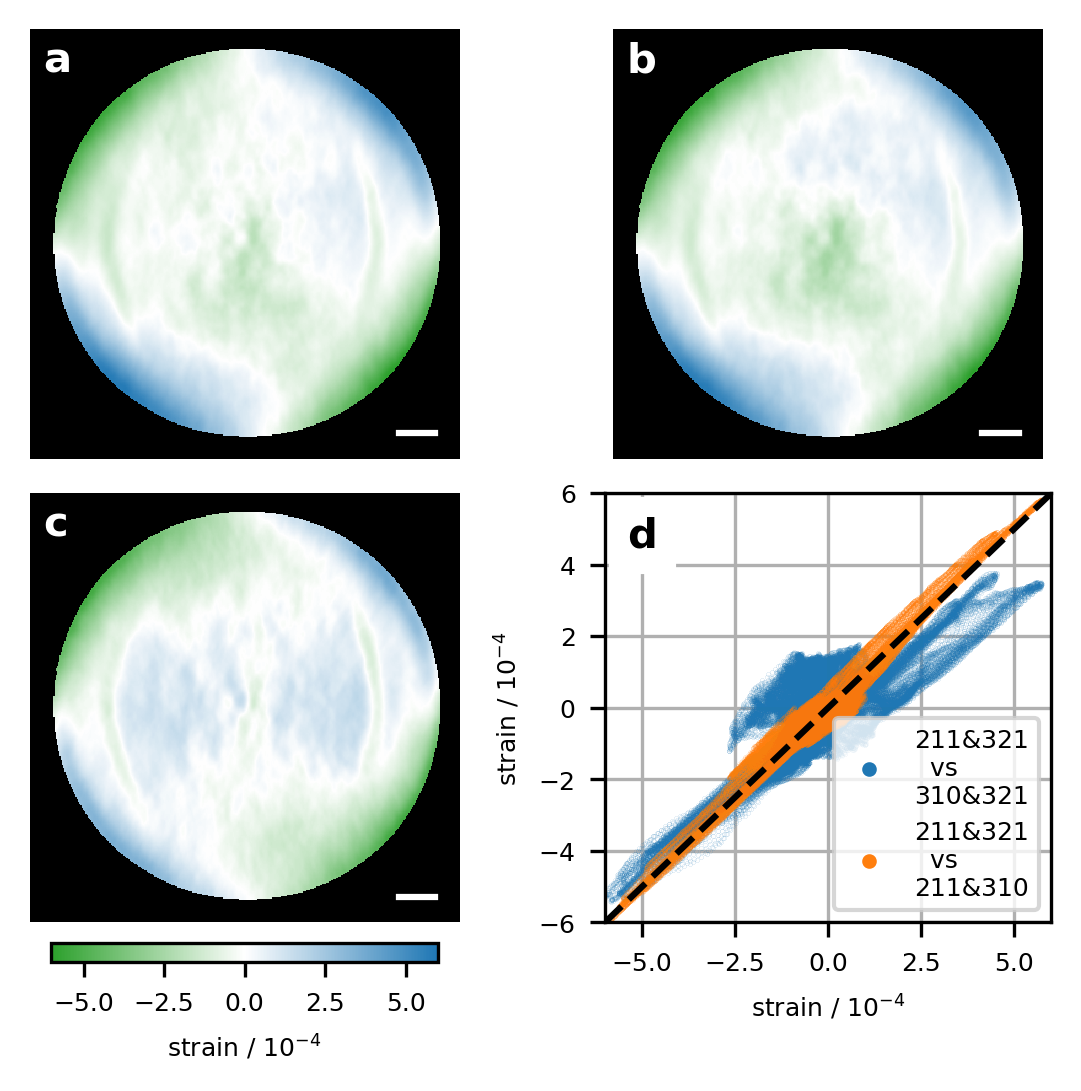}
    \caption{Comparison of $\epsilon_{11}$ reconstructions from different pairs of diffraction rings. (a) 211 \& 310 pair. (b) 211 \& 321 pair. (c) 310 \& 321 pair. Differences are found mainly in the center of the sample at low strain values. (d) Scatter plot of reconstruction values of the 211 \& 310 (orange) and 310 \& 321 pair (blue) versus the 211 \& 321 pair. Corresponding pair-wise correlation coefficients are $r=0.985$, $r=0.862$ and $r=0.832$, respectively.}
    \label{}
\end{figure}

%%%% comparison with korsunsky %%%%%

%%%% stress tensor components %%%%%
\begin{figure}[htbp]
    \centering
    \includegraphics[width=0.5\linewidth]{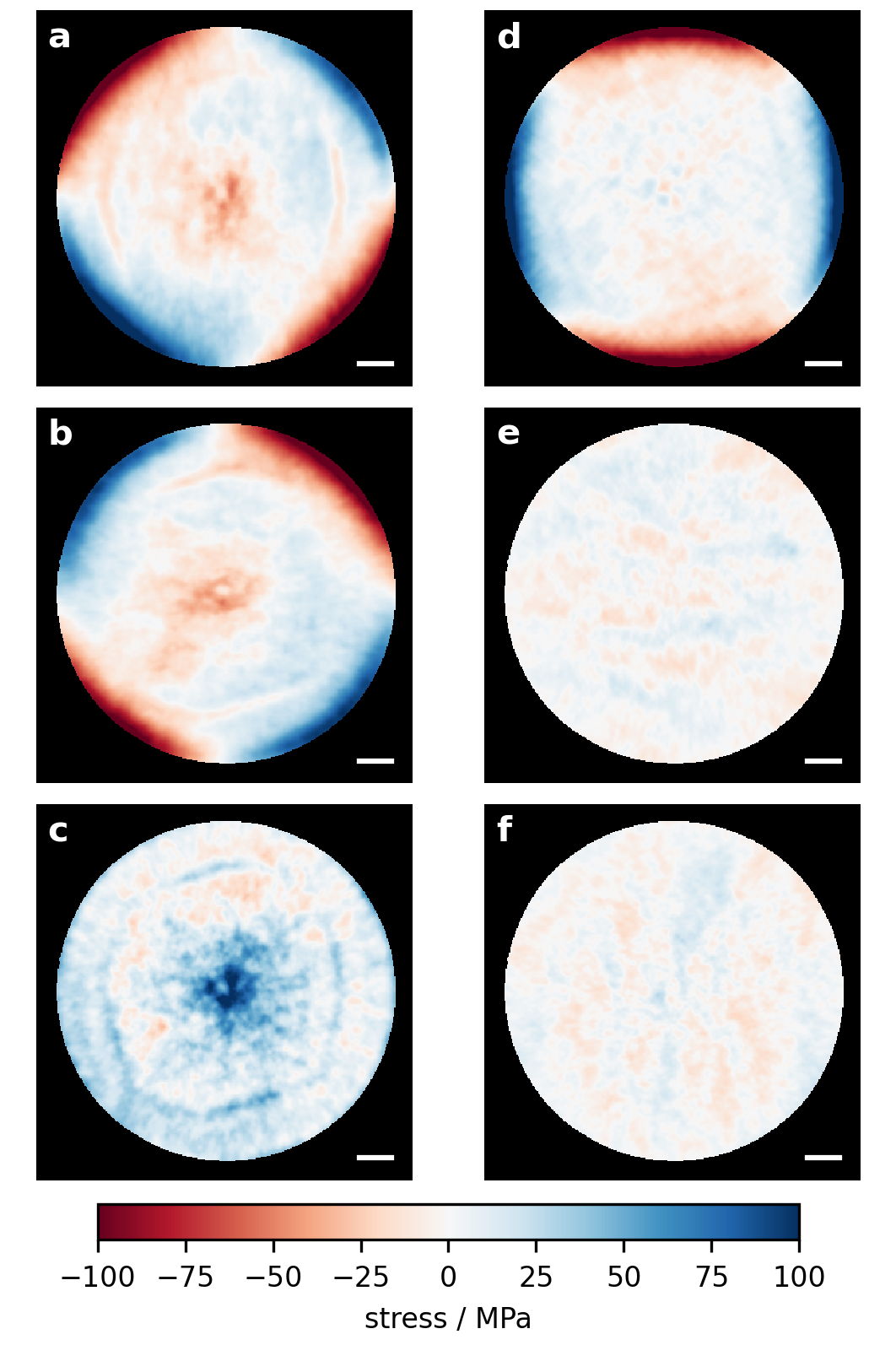}
    \caption{Stress tensor component maps of a spring steel rod. (a) $\sigma_{11}$, (b) $\sigma_{22}$, (c) $\sigma_{33}$, (d) $\sigma_{12}$,
    (e) $\sigma_{13}$, (f) $\sigma_{23}$. Scale bars are 300~$\mu$m. These are also shown in Fig.~\ref{fig:STT_setup}b.}
    % This is 06_rod stress ringinds[2,6] lambda35 Niter 219 13095sec savinteger26380
    \label{fig:control_stresses}
\end{figure}

%%%% sigma 33 - shot peening coverage? %%%%%

% \red{Reviewers: Lionheart, Withers, Grunewald, Niendorf, Christoph Genzel (Berlin), I.C. Noyan }

\end{document}